\documentclass{elsart}

\usepackage{natbib}

\usepackage{epsfig}

\usepackage{amssymb}

\begin{document}

\begin{frontmatter}



\title{HI imaging of Galaxy Clusters at z$\approx$0.2, \\
 a Pilot Survey of Abell 963 and Abell 2192}


\author{Marc Verheijen\thanksref{1}},
\author{Jacqueline van Gorkom\thanksref{1}\thanksref{2}},
\author{Arpad Szomoru\thanksref{3}},
\author{K.S. Dwarakanath\thanksref{4}}, 
\author{Bianca Poggianti\thanksref{5}}, 
\author{David Schiminovich\thanksref{2}}

\address[1]{Kapteyn Astronomical Institute, University of Groningen
    Postbus 800, 9700 AV Groningen, The Netherlands}
\address[2]{Department of Astronomy, Columbia University, 550 W 120th
    Street, New York, NY 10027, USA}
\address[3]{Joint Institute for VLBI in Europe, Dwingeloo, The
    Netherlands}
\address[4]{Raman Research Institute, Sadashivanagar, Bangalore
    560 080, India}
\address[5]{Padova Astronomical Observatory, Vicolo
    Osservatorio 5, 35122 Padova, Italy}

\begin{abstract}
A pilot study with the new powerful backend of the Westerbork
Synthesis Radio Telescope (WSRT) of two galaxy clusters at z=0.2 has
revealed neutral hydrogen emission from 39 galaxies. At these
redshifts, the WSRT provides an instantaneous velocity coverage of
18,054 km/s. The volume probed for each cluster is $1.7 \times 10^4$
Mpc$^3$, with spatial and velocity resolutions of 54$\times$86 kpc$^2$
and 19.7 km/s, covering both clusters and the large scale structure in
which they are embedded.

The spatial distribution of the H~I detected galaxies is very
different for the two clusters. In Abell~963, most of the gas rich
galaxies are located to the northeast, at 1$-$3 Mpc from the
cluster center in projection. Their velocities are slightly redshifted
with respect to the cluster mean. This could be a gas rich group
falling in from the front.  Abell~2192 is less massive and more
diffuse, with the gas rich galaxies more uniformly spread over a
large region around the cluster. The H~I masses of the detected
galaxies range from 5$\times$10$^9$ to 4$\times$10$^{10}$
M$_{\odot}$. Some H~I rich galaxies are spatially resolved and
rudimentary rotation curves are derived, showing the prospect for
Tully-Fisher studies of different galaxy populations in these
environments.

Only one galaxy is detected within a 1 Mpc radius from the center of
the Butcher-Oemler cluster Abell 963, and none of the blue B-O
galaxies which are all located within the central Mpc. The H~I
detected galaxies outside the central Mpc are of similar colour and
magnitude as the non-detected B-O galaxies, indicating that the blue
B-O galaxies are gas poor compared to their counterparts in the field.

\end{abstract}


\end{frontmatter}


\section{Introduction}
\label{}

The morphological mix of galaxy types is very different in the centers
of clusters than in the field. Ellipticals and S0's dominate in dense
clusters, spirals and irregulars dominate elsewhere.  Recent studies
have shown that the galaxy population in clusters at intermediate
redshifts evolves over relatively short timescales. Beyond
z$\approx$0.2, clusters have a larger fraction of blue galaxies,
indicative of ongoing star formation; the so called Butcher-Oemler
(B-O) effect (Butcher and Oemler, 1978).  The morphological mix in
clusters also changes with redshift. Although the fraction of
ellipticals remains unchanged from z=1 to 0, clusters at z$\approx$0.5
have a significant fraction of spirals and hardly any S0's, while at
z$\approx$0.2 the situation is reversed (Dressler et al 1997, Fasano
et al 2000, Lubin, Oke \& Postman 2002). Recent data from the local
universe, however, suggest that the changes occur more
gradually. There are smooth gradients in star formation rate, gas
content and morphological mix, out to several Mpc from the cluster
centers (Goto et al 2003; Balogh et al 1998; Solanes et al
2001). There is an ongoing debate whether it is the cluster
environment that drives the morphological evolution of galaxies or
whether it is the field population, that continues to accrete onto
clusters, that evolves with redshift.

Despite all these data, nothing is known about the actual {\bf gas}
content of cluster galaxies beyond z=0.08. The gas content is a
critical parameter in environmentally driven galaxy evolution, since
gas is the fuel for star formation. The main impediments to obtaining
data on the H~I content of galaxies at larger redshifts are the
necessarily long integration times and the occurrence of man-made
interference at those frequencies. So far, H~I emission has been
detected from only one galaxy at z=0.176 (Zwaan et al 2001).

However, if one samples a large volume and observes many galaxies at
once, it would be worth the long integration. Radio synthesis
telescopes, with their large field of view and high angular
resolution, are the ideal instruments for this. One drawback of
synthesis arrays has been their limited instantaneous velocity
coverage, insufficient to cover the range of velocities usually seen
in clusters (3000-5000 km/s). This has now changed with the new
backend of the WSRT. In a single pointing, one can cover up to 18,000
km/s with sufficient resolution, probing the entire velocity range of
the cluster and a very significant volume in front of and behind
it. Such an observation would provide the much needed gas content of
all cluster galaxies and simultaneously observe a control sample from
the field. We have performed a pilot study for just such a survey.

We selected two clusters, known to be very different in their
dynamical state and star formation properties.  {\bf Abell 963}, at
z=0.206, has a velocity dispersion of 1350 km/s and is one of the
nearest B-O clusters with a high fraction (19\%) of blue galaxies
(Butcher et al 1983).  This massive lensing cluster is among X-ray
selected clusters currently being studied at many wavelengths.  A963
is unusually relaxed with less than 5\% substructure (Smith et al
2005).  {\bf Abell 2192}, at z=0.188, has a velocity dispersion of 650
km/s and is more diffuse with significant substructure. A2192 has
not been detected in X-rays.

\begin{figure}[t]
\begin{center}
\epsfig{file=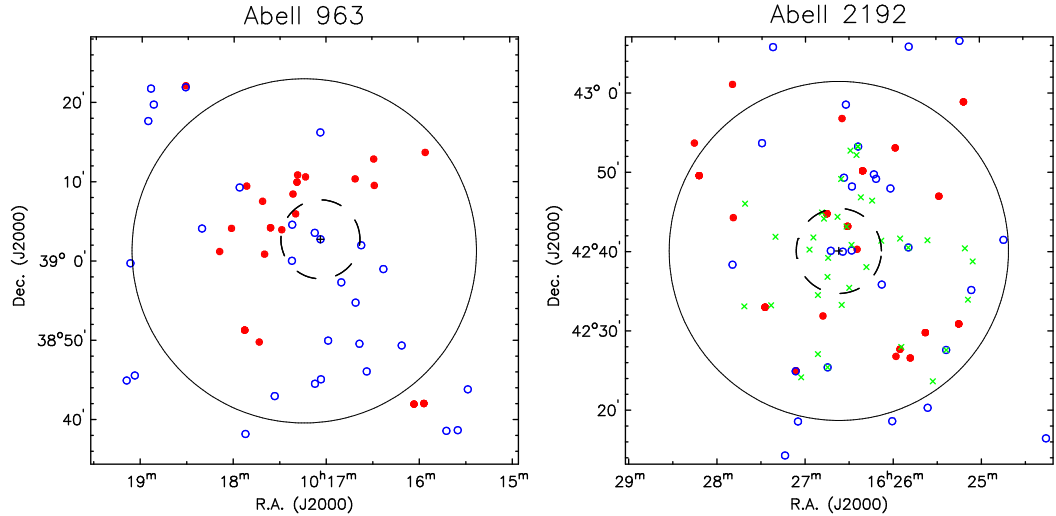,width=\textwidth}
\caption{Sky distributions of H~I detected galaxies in A963 and A2192
(filled circles). Open circles indicate galaxies with optical SDSS-DR5
redshifts within the WSRT bandpass. Crosses (A2192 only) indicate
galaxies with optical redshifts from our own observations. Large solid
circles show the FWHM of the WSRT primary beam. Small dashed
circles have a radius of 1 Mpc and are centered on the cluster cores.}
\end{center}
\end{figure}

\section{Observations}
\label{}

\begin{figure}[t]
\begin{center}
\epsfig{file=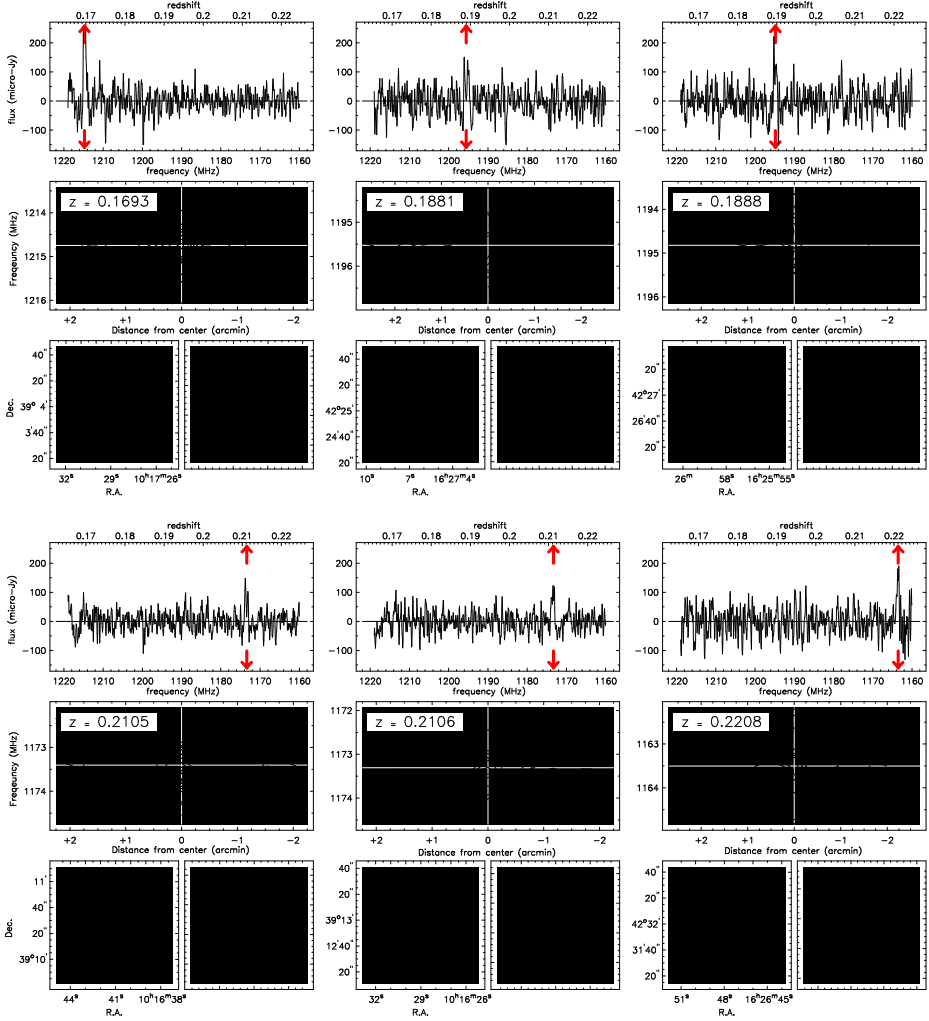,width=12.5cm}
\caption{Examples of H~I detected galaxies. Upper panels: global HI
profiles over the full frequency range. Middle panels:
position-velocity diagrams. Horizontal lines indicate systemic
velocities, vertical lines correspond to the galaxy centers. Note that
galaxies are spatially and kinematically resolved. Bottom left panels:
integrated HI maps (contours) on top of an optical image. Bottom right
panels: blow-up of the optical image. Note that several galaxies are
morphologically disturbed.}
\end{center}
\end{figure}

Abell 963 was observed for 20$\times$12 hrs in February 2005, and
Abell 2192 was observed for 15$\times$12 hrs in July 2005, each with a
single pointing. The correlator was configured to give eight
partially overlapping 10 MHz bands (IVC's) with 256 channels per band
and 2 polarizations per channel. The frequency range covered is
1220$-$1160 MHz. The main challenge in the data processing is the
presence of low level radio frequency interference and we used a
combination of visual inspection and various filtering algorithms to
remove bad data. To calculate the bandpasses of the IVC's, we used
narrow median window filtered versions of the spectra of the flux
calibrators, and applied these corrections to the cluster data. A
linear fit in frequency was then made to the UV data and data
deviating by more than 8$\sigma$ were clipped. The data from all 8
IVCs were combined into a single datacube for each cluster, comprising
1600 channels. The typical rms noise levels per frequency channel are
68 $\mu$Jy/beam at 1178 MHz for A963, and 91 $\mu$Jy/beam at 1196 MHz
for A2192. After Hanning smoothing, the restframe velocity resolution
is 19.7 km/s while the synthesized beam is 17$\times$27 arcsec$^2$ at
1190 MHz.

\begin{figure}[t]
\begin{center}
\epsfig{file=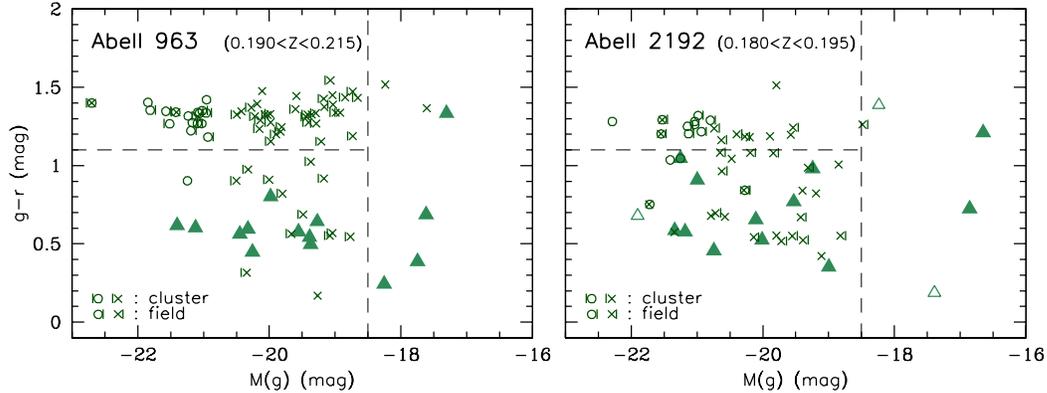,width=\textwidth}
\caption{SDSS-based Colour-Magnitude diagrams for galaxies with known
redshifts near those of the clusters. Triangles: H~I detected
galaxies. Open circles: galaxies with redshifts in SDSS-DR5. Crosses:
for A963, galaxies in the cluster core with optical redshifts from
Lavery \& Henry (1986), not included in Figure 1, and for A2192,
galaxies with optical redshifts from our observations. Horizontal
dashed lines separate the red sequence from the blue cloud. Vertical
dashed lines indicate the luminosity of a LMC-like galaxy. Symbols
with vertical bars next to them correspond to galaxies used to stack
their H~I spectra as shown in Figure 4.}
\end{center}
\end{figure}

In search of H~I emission, we smoothed the data to a restframe
velocity resolution of 40 km/s and visually inspected both
datacubes. We have identified 20 H~I detections in the field of A963,
and 30 in that of A2192. Subsequently, we searched the Sloan Digital
Sky Survey for optical counterparts within the synthesized beam and
found 17 objects for A963 and 22 for A2192. We consider these 39 H~I
detections as secure and the remaining 11 as tentative.  Based on the
local H~I mass function (Zwaan et al 2003), the surveyed volume, the
achieved noise levels, and a homogeneous distribution of galaxies, the
predicted number of detected galaxies is 22 for A963 and 17 for
A2192. Therefore, the observed detection rate in our pilot data is as
expected.

\section{Results}
\label{}

Figure 1 shows the distributions of H~I detected galaxies on the sky
(filled circles) which are very different for the two clusters. The
gas rich galaxies in the field of A963 are strongly clustered toward
the northeast; many gas-rich galaxies are within a projected distance
of 3 Mpc from the cluster center, but only one galaxy is detected
within 1 Mpc. In the field of A2192, the gas rich galaxies are more
uniformly distributed over the surveyed area. This difference in the
spatial distributions is to some extent also reflected in the velocity
distribution of the H~I detected galaxies. The H~I rich galaxies in
A963 are more strongly clustered in velocity, and slightly redshifted
with respect to the cluster mean.  Most likely, this is a group of gas
rich galaxies falling into A963 from the front.

Figure 2 shows examples of data on individual galaxies at increasing
redshifts. The individual H~I detections have H~I masses between
5$\times$10$^9$, close to our detection limit, and 4$\times$10$^{10}$
M$_{\odot}$. Depicted are a small group of interacting galaxies in the
foreground of A963 (upper left), several spatially resolved galaxies
at the redshifts of the clusters, and a background galaxy in the field
of A2192 (lower right). The latter represents the highest redshift H~I
emission detected to date.  Although the synthesized beam is rather
large, the galaxies are clearly resolved both kinematically and
spatially. The position velocity diagrams show rudimentary rotation
curves and look promising for future Tully-Fisher studies of galaxies
in different environments. Note that some of the optical images show
disturbed galaxies and several seem to have close companions.

Figure 3 shows SDSS-based colour-magnitude diagrams for all galaxies
with various optical (circles and crosses) and HI (triangles)
redshifts within the velocity range of the clusters.  For A963, the
red sequence and blue cloud can be clearly discerned, but for A2192
the red sequence is less obvious. The optical magnitudes and colours of
the H~I detected galaxies are similar to those of the blue galaxies
with optical redshifts in both fields. Although none
of the blue galaxies in the center of A963 are detected in H~I, those
that are detected in H~I outside the centers have similar magnitudes
and colours as the blue B-O galaxies. This suggests that it is the
location of the blue B-O galaxies that affects the H~I detection rate.

We can probe even deeper by stacking the H~I spectra of galaxies with
known optical redshifts.  This is done separately for the blue
and red galaxies of Lavery and Henry (1986) in the central Mpc of
A963, and for the red and blue galaxies in both fields. The stacked
spectra are shown in Figure 4.  Even in the stacked spectra there is
no statistical detection of the blue B-O galaxies in the central Mpc
of A963. In contrast, the blue galaxies in the fields outside the
central Mpc do show a statistical detection in the stacked H~I
spectra.

\begin{figure}[t]
\begin{center}
\epsfig{file=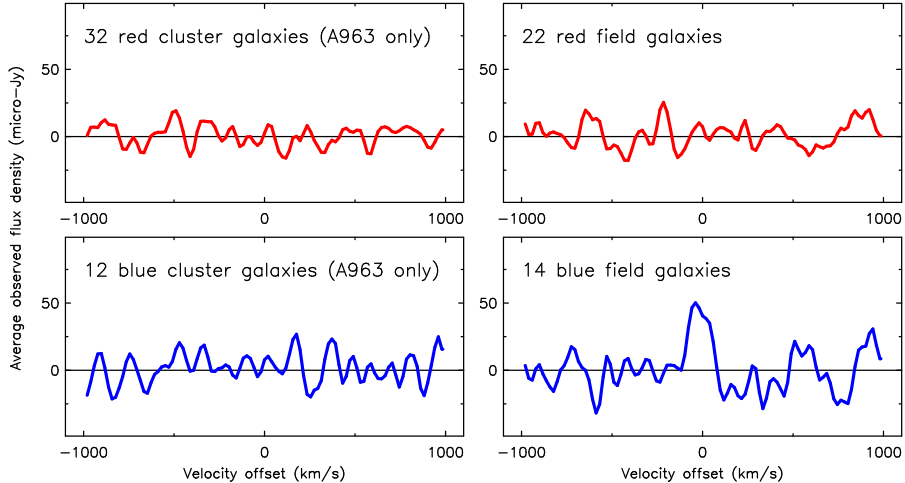,width=12cm}
\caption{Stacked H~I spectra of red and blue galaxies with optical
redshifts in the cluster cores ($<$1Mpc) and the surrounding
field. The blue B-O galaxies in the core of A963 are not detected
statistically while similar blue galaxies in the field are clearly
detected with an average H~I mass of 2$\times$10$^9$ M$_\odot$.}
\end{center}
\end{figure}

\section{Discussion}
\label{}

We present the results of a pilot study to demonstrate the feasibility
of an H~I emission line survey at z=0.2. Our detection rate and
achieved noise levels show that it is now entirely feasible to do deep
searches for H~I in emission at cosmologically interesting
distances. While our detection rate is similar for both clusters, the
distribution of the H~I detections is more highly clustered for the
dense B-O cluster. Most noticeably, in A963 we detect 17 gas rich
galaxies outside a radius of 1 Mpc from the cluster center, and only
one just within that radius. At the same time we know that A963 has a
significant fraction of blue galaxies in its central region. Our
stacked H~I spectra suggest that the blue B-O galaxies have
significantly smaller H~I masses (on average) than similar blue
galaxies outside the central Mpc.  Since the H~I detected galaxies are
of similar colour and magnitude as the non-detected B-O galaxies in
the central region, we conclude that it is the location that matters.
These results are at odds with recent claims that the B-O effect
predominantly occurs in subclusters and groups (Kodama et al 2001,
Tran et al 2005).  Although we can not rule out projection effects,
the most straight forward interpretation of our results is that the
blue B-O galaxies have lost (a significant fraction of) their gas when
they came to within 1 Mpc of the center of A963.

Our final survey will probe a huge volume around each cluster down to
an H~I mass limit of 8$\times$10$^8$ M$_{\odot}$. We expect to detect
hundreds of galaxies in the clusters and the large scale structure in
which they are embedded. This will give the first optically unbiased
H~I survey at z=0.2 with enough sensitivity to detect even galaxies
like the LMC in a volume similar to that of the entire local Universe
out to 25 Mpc.

The WSRT is operated by the ASTRON with support from the Netherlands
Foundation for Scientific Research (NWO). This work was partially
supported by an NSF grant to Columbia University. We are grateful to
R. Lavery for providing unpublished redshifts of galaxies in A963. The
full SDSS acknowledgement can be found at http://www.sdss.org.




\end{document}